\RequirePackage{fix-cm}
\documentclass[epj,twocolumn]{svjour}
\usepackage{graphicx}
\usepackage{epstopdf}
\usepackage{booktabs}
\usepackage{mathrsfs} 
\graphicspath{{fig/}}
\usepackage{mathtools}
\usepackage{url}
\usepackage{float}
\usepackage{tikz,url,float}
\usepackage{dcolumn}
\usepackage{bm}
\usepackage{color}
\usepackage{cite}
\usepackage{placeins}
\usepackage[colorlinks=true,linkcolor=blue,citecolor=blue]{hyperref}
\input{symdef.sty}

\journalname{Eur. Phys. J. E}

\begin{document}
\title{Pseudo-turbulence in two-dimensional buoyancy driven bubbly flows: a DNS study}

\author{Rashmi Ramadugu,Vikash Pandey, and Prasad Perlekar}

\institute{TIFR Center for Interdisciplinary Sciences, Hyderabad, 500107, India \\
              \email{rashmir@tifrh.res.in}           
}

\date{\today}

\authorrunning{Ramadugu. R, Pandey. V, Perlekar. P }
\titlerunning{Pseudo-turbulence in buoyancy driven bubbly flows}

\abstract{
We present a direct numerical simulation (DNS) study of  buoyancy driven bubbly flows in
two-dimensions. We employ volume of fluid (VOF) method to track the bubble interface. To
investigate spectral properties of the flow, we derive the scale-by-scale energy budget
equation. We show that the Galilei  number ($\Ga$) controls different scaling regimes in the energy spectrum.  For high Galilei numbers, we find the presence of an inverse energy cascade. Our study indicates that the density ratio of the bubble with the ambient fluid or the presence of coalescence between the bubbles does not alter the scaling behaviour.}

\PACS{{47.55.D-}{Drops and bubbles}}
\maketitle

\section{Introduction}
A swarm of bubbles rising under gravity generates complex spatio-temporal flow patterns, often referred to as pseudo-turbulence (PT) or bubble induced agitation. Although the trajectory and wake of an isolated bubble depend on its viscosity and density contrast with the surrounding fluid  \cite{clift1978, bha81,kel97,wang2014experimental,tri15,fil15}, the statistical properties of the flow generated by the bubble swarm are found to be universal \cite{mudde_rev_2005,martinez_2010,riss18,mat20,pan20}. A key feature of PT is the power-law scaling in the energy spectrum with an exponent of $-3$ either in frequency $f$ or the wave-number $k$ space \cite{martinez_2010,risso_legendre_2010,mendez,risso_billet_2014},  explained by the balance of energy production by wakes with viscous dissipation \cite{lance_1991,riss18,pan20}. 

The key non-dimensional numbers that  characterise pseudo-turbulence are the Bond number
$\Bo\equiv \delta \rho g d^2/\sigma$ (ratio of
the buoyancy forces to the surface tension forces), the  Galilei number $\Ga \equiv
\sqrt{\rho_f \delta \rho g d^3}/\mu$ (ratio of the buoyancy forces to the viscous forces),
and the Atwood number $\At \equiv \delta \rho/(\rho_f+\rho_b)$, where $\rho_f$ is the density of the ambient fluid, $\rho_b$ is the bubble density, $\delta \rho\equiv \rho_f-\rho_b$, $g$ is the acceleration due to gravity, and $d$ is the initial bubble diameter.

Experiments in three-dimensions (3D) at low volume fraction $\phi \approx 2\%$ observe the $k^{-3}$ scaling in the energy spectrum both within and in the wake of the bubble swarm for $202 \leq\Ga \leq 396$ \cite{risso_legendre_2010}. In Hele-Shaw geometry, experiments \cite{risso_rising_2012,risso_billet_2014} at moderate volume fraction $\phi \approx 5-16 \%$ and $  650 \leq \Ga \leq1100$ also observe the $k^{-3}$ scaling.  

Most numerical studies have investigated PT in 3D
\cite{roghair, bunner_tryg_2002,pan20}  and found that the statistical properties of the flow to
be universal for a wide range of Atwood $\At$, Bond $\Bo$, and Galilei $\Ga$ numbers. Using a
scale-by-scale energy budget analysis, Ref.~\cite{pan20} showed that both the surface tension
and kinetic energy flux contribute to the net energy production at scales smaller than $d$. 

In comparison to 3D, there have been very few studies on two-dimensional (2D) bubbly flows. Early numerical simulations at low $\Ga=5.4$ \cite{Tryg1996} indicated the presence of a $k^{-3}$ scaling regime. Recent high-resolution direct numerical simulation (DNS) in 2D at high $\Ga$ show the presence of a $k^{-3}$ scaling for scales smaller than the bubble diameter and an inverse energy cascade for larger scales both within \cite{per18} and in the wake of a swarm \cite{inno20}.

Turbulence in two-dimensions is fundamentally different from its counterpart
in three-dimensions; in the inviscid limit ($\nu \to 0$), conservation of kinetic energy leads
to an inverse energy cascade from forcing scales to large-scales and a direct enstrophy cascade from forcing scales to small scales~\cite{fjo53,kraic67,lei67,bat69,eyink1996,per09,Kraichnan1980,boffetta,pan17}. On the other hand, only a forward energy cascade is possible in three-dimensions because of finite energy dissipation even when viscosity tends to zero~\cite{frisch,pope,per09}.\\
Several numerical and experimental studies of homogeneous, isotropic turbulence in two- and three-dimensions have studied and verified the proposed energy transfer mechanisms and scaling laws in real and spectral space \cite{frisch,pope,ant97,per09,boffetta,bif15,pan17}. Scale-by-scale energy budget analysis provides a natural way to investigate the interaction between different length scales.\\
In this paper, we present a DNS study in 2D to investigate pseudo-turbulence generated by buoyancy-driven bubbly flows for small and large Ga, At, and Bo numbers. We characterize the flow in terms of the bubble size distribution,  energy spectrum, and use the scale-by-scale energy budget analysis to study energy transfer mechanisms. Below we summarize the main results of our DNS study:
\begin{enumerate}
\item The average bubble diameter $D \sim \Bo^{-1/2}$ in a bubbly flow.
\item  The PT $k^{-3}$ scaling  in the energy spectrum appears for scales larger than the bubble diameter for small $\Ga$, whereas it appears for scales smaller than the bubble diameter for large $\Ga$.  
\item For large $\Ga$, we observe the presence of an inverse energy cascade and a $k^{-5/3}$ scaling in the energy spectrum for scales larger than the bubble diameter. 
\item Our scale-by-scale energy budget analysis for large $\Ga$ reveals: $(i)$ a negative energy flux for scales larger than the bubble diameter, and  $(ii)$ the net energy production balances viscous dissipation for scales smaller than the bubble diameter.
\end{enumerate}

\begin{table*}[ht!]
     \caption{\label{tab:runs} The parameters $N$,$L$, $d$, $\Ga$, $N_b$, $\Bo$, $\phi$, $\At$,
     the viscous dissipation $\epsilon_\mu$, and the energy injected due to buoyancy
 $\epsilon_{inj}$ for non-Boussinesq ($\tt{NB}$) and Boussinesq ($\tt{B}$) runs. We use
 front-tracking method \cite{Tryg2001}  for the Boussinesq runs where coalescence is arrested.
 $N_b$ represent initial number of bubbles.  We choose $\mu_f/\mu_b = 1$ and  $g=1$  for all the
 runs except $\tt{NB1}$ - $\tt{NB5}$ where $\mu_f/\mu_b = 20$.  All the simulations were
 conducted at TIFR-H Kohinoor3 cluster \cite{facility}. The high $\Ga$ runs took $\approx 5$ hours for 1$\tau_s$ on 64 CPUs.}

 \begin{center}
\begin{tabular}{ccccccccccccc}
 \hline\noalign{\smallskip}
          
       $runs$   & $L$   &$d$ & $N$  &$ \rho_f$   & $\Ga$    & $N_b$ & $\Bo$      & $\phi$ & $\At$ &
       $\epsilon_\mu$  & $\epsilon_{inj}$ & $\epsilon_w$ \\ 
       \noalign{\smallskip}\hline\noalign{\smallskip}
       
       $\tt{NB1}$    & $12^2$   & 0.4 & $1024^2$ &25.0 & 2.7    & 144    & 1.0       & 0.12   & 0.9 & 
       $2.1 \pm 0.6$  & $1.8 \pm 0.1$ & $1.5$ \\
        
        $\tt{NB2}$    & $12^2$   & 0.4 & $1024^2$ &25.0 & 5.4     & 144    & 0.5        & 0.12   & 0.9 & 
       $1.3 \pm 0.7$  & $0.9 \pm 0.1$  & $1.6$ \\
        
                 $\tt{NB3}$    & $12^2$   & 0.4 & $1024^2$& 25.0  & 5.4     & 144    & 1.0        & 0.12   & 0.9 & 
       $1.9 \pm 0.1$  & $1.8 \pm 0.1$ & $1.5$\\
        
         $\tt{NB4}$    & $12^2$   & 0.4 & $1024^2$ &25.0 & 5.4     & 144    & 2.0       & 0.12   & 0.9& 
       $0.9 \pm 0.1$  & $1.3 \pm 0.2$ & $ 1.2$\\

         $\tt{NB5}$    & $12^2$   & 0.4 & $1024^2$ &25.0  & 16    & 144    & 1.0        & 0.12   & 0.9 & 
       $1.2 \pm 0.1$  & $1.4 \pm 0.1$ & $1.5$\\

        $\tt{NB6}$    & $512^2$   & 20.0 & $2048^2$  &1.12& 312     & 144    & 0.5        & 0.17   & 0.08 & 
       $(2.8 \pm 0.7)\cdot10^{-2}$  & $(3.4 \pm 0.2)\cdot10^{-2}$ & $3.9\cdot 10^{-2}$\\
       
        $\tt{NB7}$    & $512^2$   & 20.0 & $2048^2$  & 1.12 & 312     & 144    & 1.0        & 0.17   & 0.08 & 
       $(2.8\pm 0.3)\cdot10^{-2}$  & $(3.1\pm 0.6)\cdot10^{-2}$ & $3.3\cdot 10^{-2}$\\
       
        $\tt{NB8}$    & $512^2$   & 20.0 & $2048^2$ &1.12 & 312     & 144    & 5.0        & 0.17   & 0.08 & 
       $(1.8\pm 0.2)\cdot10^{-2}$  & $(2.0\pm 0.1)\cdot10^{-2}$ & $2.2\cdot 10^{-2}$\\
     
              $\tt{NB9}$    & $512^2$   & 25.0 & $2048^2$ &1.08 & 312     & 100    & 1.0         & 0.19   & 0.08 & 
       $(3.2\pm 0.1)\cdot10^{-2}$  & $(3.2\pm 0.2)\cdot10^{-2}$ & $4.0\cdot 10^{-2}$\\
       
        $\tt{NB10}$    & $512^2$   & 20.0 & $2048^2$ &1.12 & 723     & 144    & 0.5         & 0.17   & 0.75 & 
       $(3.3\pm 0.2)\cdot10^{-1}$  & $(4.6\pm 0.5 )\cdot10^{-1}$ & $5.2 \cdot 10^{-1}$\\
        
            $\tt{NB11}$    & $512^2$   & 20.0 & $2048^2$  &1.12 & 723       & 144    & 1.0         & 0.17   & 0.75 & 
       $(3.0 \pm 0.1)\cdot10^{-1}$  & $(3.9 \pm 0.5)\cdot10^{-1}$ & $ 4.6\cdot 10^{-1}$\\
       
        $\tt{NB12}$    & $512^2$   & 20.0 & $2048^2$ & 1.12  & 723    & 144    & 10.0         & 0.17   & 0.75 & 
       $(2.7\pm 0.8)\cdot10^{-1}$  & $(3.4\pm 0.6)\cdot10^{-1}$ &  $ 2.5 \cdot 10^{-1}$\\
       
       $\tt{B1}$     & $512^2$   & 25.0 & $4096^2$   & 1.0 & 312     & 100    & 0.5         & 0.19   & 0.08 & 
       $(3.5\pm 0.3)\cdot10^{-2}$  & $(3.3\pm 0.3)\cdot10^{-2}$ & $3.0\cdot 10^{-2}$\\
\noalign{\smallskip}\hline
     \end{tabular}
     \end{center}
\end{table*}

The rest of this paper is organised as follows. In Section.~\ref{mod}, we present the governing equations and the details of our DNS. In Section~\ref{res}, we discuss our results on buoyancy driven bubbly flows. We present our conclusions in Section~\ref{concl}.

\section{Model and Numerical Details \label{mod}}
We study the dynamics of bubbly flow by using  Navier-Stokes (NS) equations with a surface tension force because of bubbles

\begin{subequations}
	\begin{eqnarray}
		D_t c &=& 0, ~\text{and}\hspace{2mm} \nabla\bm{\cdot}\bm{u} = 0,\\
	\rho(c) D_t\bm{u} &=&   \nabla \bm{\cdot} [2 \mu(c) {\cal S}]  - \nabla p 
	 + {\bm F}^\sigma  + {\bm F}^g. \label{eqn:mom} 
	\end{eqnarray}
	\label{eqn:ns}
\end{subequations}

Here, $D_t = \partial_t + (\bm{u}\cdot\nabla)$ is the material derivative, $c$ is an indicator
function whose value is $0$ inside the bubble phase and $1$ in the fluid phase. ${\bm
F}^{\sigma} \equiv \sigma \kappa \nabla c$ is the force because of the surface tension
\cite{brackbill_1992,Basilisk}, ${\bm F}^{g} \equiv  [\rho_a-\rho(c)] g\hat{\bm{e}}_y$ is the
buoyancy force, ${\bm u} = (u_x,u_y)$ is the hydrodynamic velocity, $p$ is the pressure, the
local density $\rho(c)\equiv \rho_f c + \rho_b (1-c)$, the local viscosity $\mu(c)\equiv \mu_f c
+ \mu_b (1-c)$, $\rho_b$ ($\rho_f$)    is  the bubble (fluid) density, $\mu_b$ ($\mu_f$) is the
bubble (fluid) viscosity, $\phi \equiv
[\int (1-c) d{\bm x}]/L^2$ is the bubble volume fraction, ${\cal S}\equiv (\nabla {\bm u} +
\nabla {\bm u}^T)/2$ is the rate of deformation tensor, $\sigma$ is the coefficient of surface
tension and $\kappa$ is the curvature. For small Atwood numbers, \Eq{eqn:mom} can be further
simplified by invoking Boussinesq approximation whereby, $\rho(c)$ in the left-hand-side of
\Eq{eqn:mom} is replaced by the average density $\rho_a = \overline{\rho(c)}\approx
(\rho_f+\rho_b)/2$, where the $\overline{(.)}$ denotes spatial averaging.   

We use a periodic box of volume $L^2$ and discretize it with $N^2$ collocation points. We initialize the simulation with a quiescent flow-field  ${\bm u}({\bm x},t=0)=0$ and place $N_b$ bubbles at random positions well-separated from each other. We numerically integrate \Eq{eqn:ns} using a second-order accurate  volume of fluid (VOF) solver Basilisk \cite{Basilisk,popinet2018} which has been used to study a variety of multiphase flow problems.  For a review of various numerical schemes used for  multiphase flows and comparison between them, we refer the reader to Refs.~\cite{popinet2018,TSZ:2011,b_mulp_flow}. The parameters that we use in our DNS are listed in Table~\ref{tab:runs}.

\section{Results \label{res}}
\begin{figure}[!h]
\begin{center}
\includegraphics[width=0.75\linewidth]{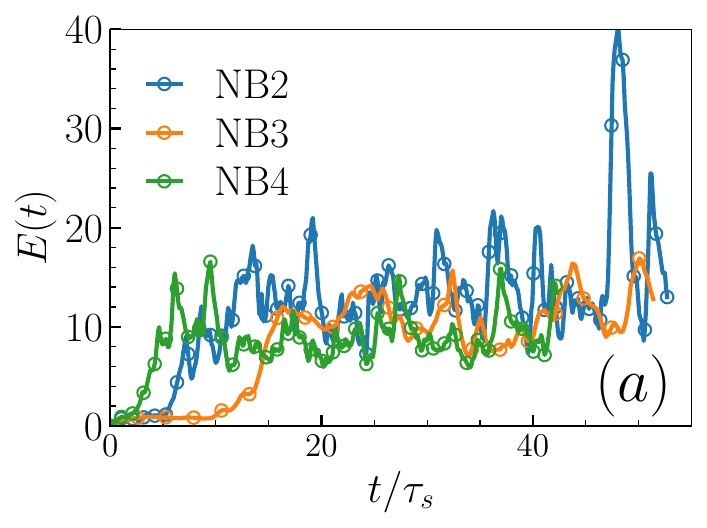}\\
\includegraphics[width=0.75\linewidth]{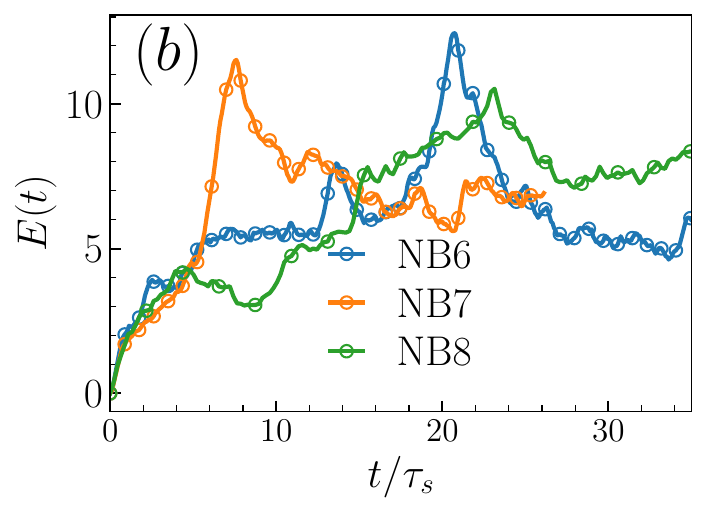}\\
\includegraphics[width=0.75\linewidth]{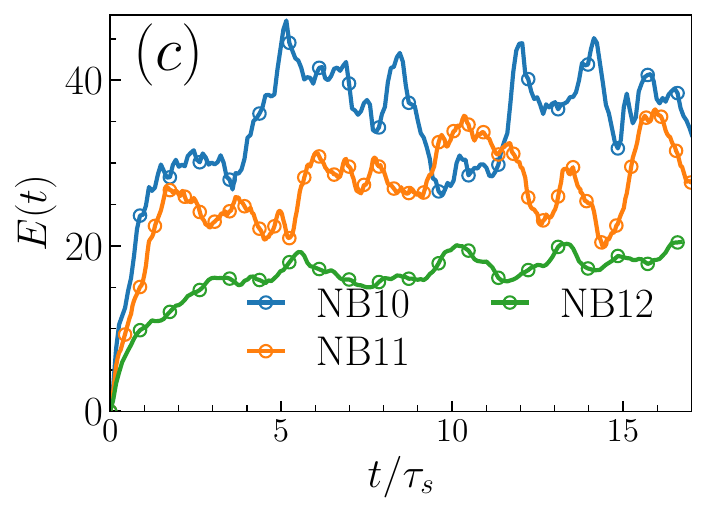}
\caption{\label{ken} Plot of kinetic energy ($E(t)$)  versus time for (a) runs {\tt NB2-NB4} ($\Ga=5.4, \At=0.9$), (b) runs {\tt NB6-NB8} ($\Ga=312, \At=0.08$), and (c) runs {\tt NB10-NB12} ($\Ga=723, \At=0.75$).}
\end{center}
\end{figure}

In  \Fig{ken} we plot  kinetic energy $E \equiv \overline{\rho u^2/2}$ versus $t/\tau_s$ for our runs ${\tt NB2-NB4}$ ($\Ga=5.4$, $\At=0.9$), ${\tt NB6-NB8}$ ($\Ga=312$, $\At=0.08$), and ${\tt NB10-NB12}$ ($\Ga=723$, $\At=0.75$), where 
$\tau_s = L/\sqrt{g d}$ is the approximate time taken by an isolated bubble to traverse the entire domain. After an initial transient, a statistically steady state is attained where the bubbles  continuously merge and break to form a stationary suspension.

\begin{figure}[!h]
\begin{center}
\includegraphics[width=0.48\linewidth]{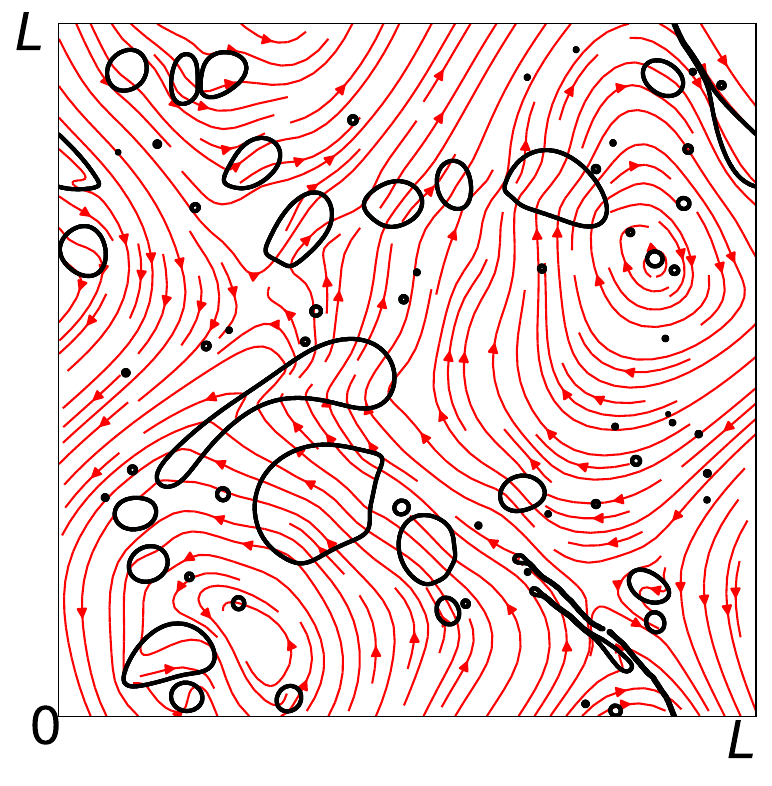}
\put(-20,20){\bf (a)}
\includegraphics[width=0.48\linewidth]{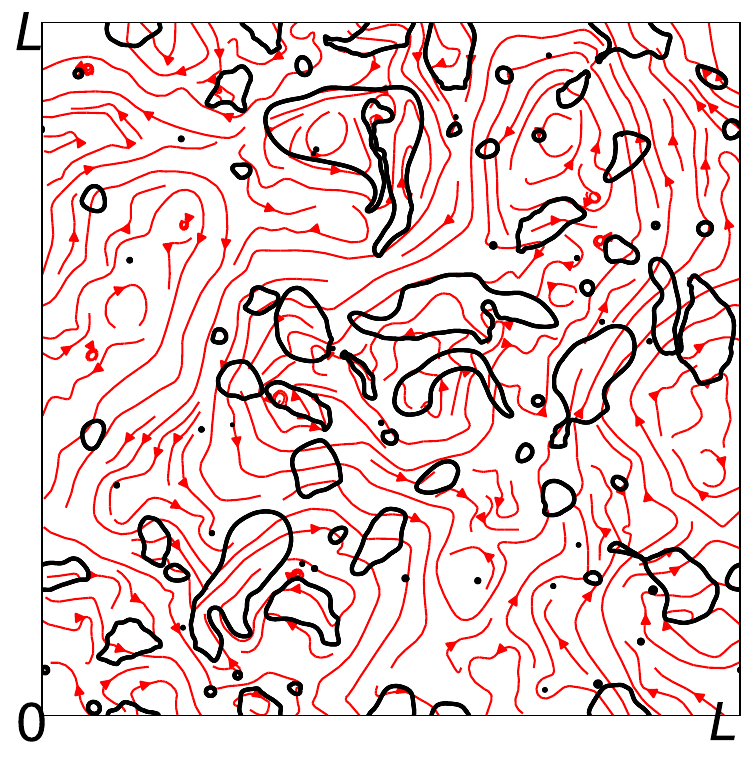}
\put(-20,20){\bf (b)}
\end{center}
\caption{\label{snap} Instantaneous velocity streamline in the steady-state with overlaid bubble positions for our runs {\tt NB3}(a), and {\tt NB11} (b).} 
\end{figure}

In \Fig{snap} we show the representative plots of the steady-state velocity streamlines overlaid
with bubble positions for $\Ga=5.4$ and $\Ga=723$.  For low $\Ga$ numbers, the typical flow
eddies are larger or comparable to the bubbles in the suspension \cite{Tryg1996}, whereas at 
high $\Ga$ numbers smaller eddies are also formed.

\subsection{Bubble size distribution}
\begin{figure*}[!ht]
\begin{center}
	\includegraphics[width=0.45\linewidth]{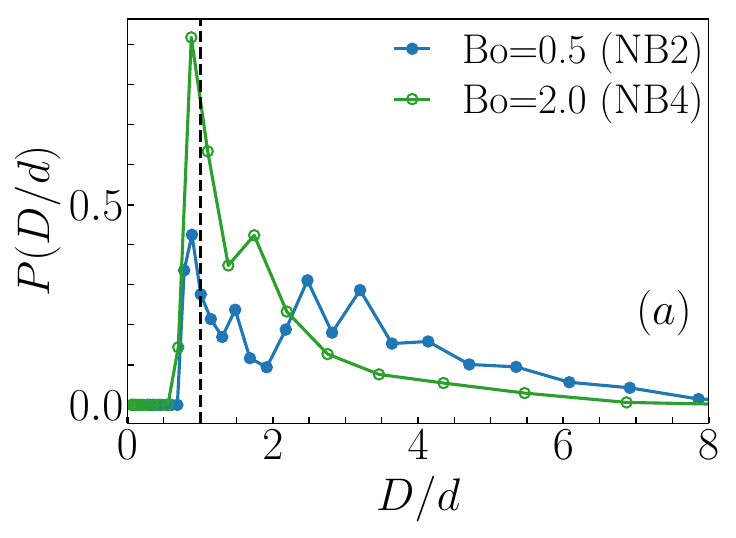}
        \includegraphics[width=0.45\linewidth]{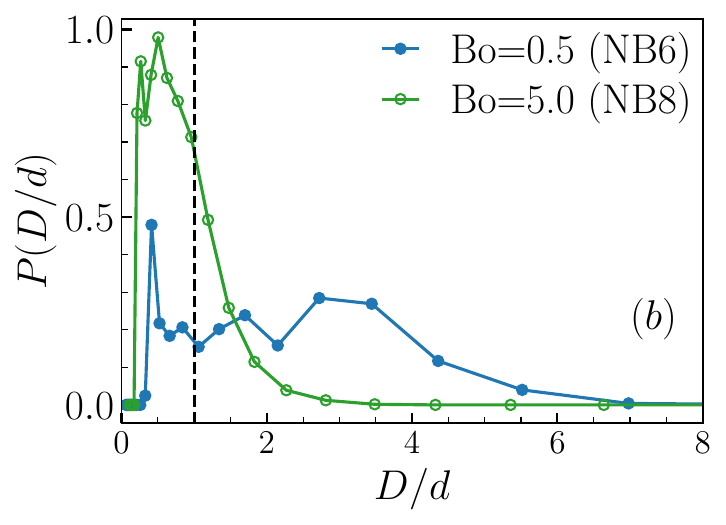}\\		
        \includegraphics[width=0.45\linewidth]{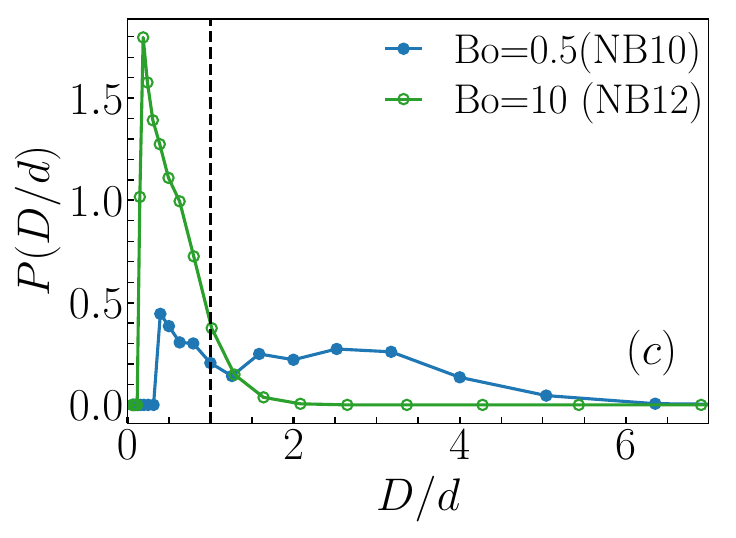}
	\includegraphics[width=0.45\linewidth]{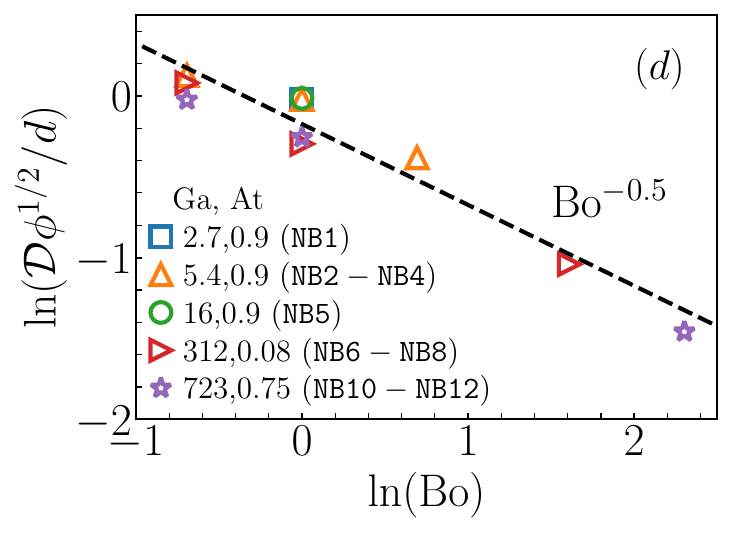}
\end{center}
\caption{\label{plot:bsd_hat}(a-c)  Pdf of the bubble diameter $P(D/d)$ versus $D/d$ for different $\Bo$ and $\Ga$ [(a) $\Ga=5.4$, (b) $\Ga=312$, and  (c) $\Ga=723$]. The vertical dashed line marks the initial bubble diameter. (d) Plot of  the average bubble diameter $D \phi^{1/2} /d$ versus  $\Bo$ for different combinations of  $\Ga$ and $\At$ (runs {\tt NB1}-{\tt NB12}).  The black line shows the theoretical prediction  \eqref{avD}. }
\end{figure*}

We track bubbles and evaluate the diameter of an equivalent circle $D_i(t) \equiv
\sqrt{4A_i(t)/\pi}$ from their area, where the subscript $i$ indicates the bubble index. In
\subfig{plot:bsd_hat}{a-c}, we plot the probability distribution function (pdf) of the bubble
diameter $P(D)$ for different values of $\Ga$ and $\Bo$. The competition between breakups and
coalescence of bubbles due to the flow determines the pdf shape. For large $\Bo>1$ (small
$\sigma$), the breakup of bubbles is more dominant than coalescence, and the pdf's peak is to the
left of the initial bubble diameter $d$. On reducing the $\Bo$ (increasing $\sigma$),
coalescence becomes more dominant than breakups, the pdf broadens, and a secondary peak starts to appear at $D>d$.

Consider a bubble whose diameter  is the same as the average bubble diameter ${\mathcal D}$ in the suspension, 
where 
\begin{equation}
\mathcal{D} = \int D P(D) dD.
\label{avf}
\end{equation}
The rise velocity of this bubble $U \sim \sqrt{\phi \delta \rho g \mathcal{D}/\rho_f}$ is
determined by the balance of buoyancy with the drag \cite{clift1978}. Whether such a rising
bubble breaks or not can be estimated by balancing the 2D bubble kinetic energy $\sim \rho_f U^2
{\mathcal D}^2$ with the bubble surface energy $\sim \sigma \mathcal{D}$ \footnote{Note that in 3D, the balance of the bubble kinetic energy $\sim \rho_f U^2 \mathcal{D}^3$ with bubble surface energy $\sim \sigma \mathcal{D}^2$ also gives \eqref{avD}.} to get, 
\begin{equation}
\mathcal{D} \sim \sqrt{\frac{\sigma}{\phi \delta \rho g}} \equiv \frac{d}{\sqrt{\phi \Bo}}.
\label{avD}
\end{equation}

In   \subfig{plot:bsd_hat}{d} we plot the average bubble diameter $\mathcal{D}$ \eqref{avf} for all the runs given in Table~\ref{tab:runs} and find it to be in good agreement with the theoretical prediction \eqref{avD}.

In the following, we  first discuss the steady-state energy budget and derive the scale-by-scale
energy budget equations in Sections~\ref{enb} and~\ref{esp}. Using these, we then investigate
the energy spectrum and dominant balances by varying $\Ga$ and $\At$ numbers.

\subsection{Kinetic energy budget\label{enb}}
Taking the dot product of  \Eq{eqn:mom} with ${\bm u}$ and then performing spatial averaging, we obtain the following equation for the evolution of the total energy  \cite{pan20}: 
\begin{equation}
\partial_t (\underbrace{\frac{1}{2}\overline{{\rho {\bm u}^2}}}_{E} + \underbrace{\overline{\sigma ds}}_{E_\sigma})
 = -\underbrace{ 2 \overline{\mu(c) \mathcal{S}:\mathcal{S}}}_{\epsilon_\mu} 
  + \underbrace{\overline{[\rho_a -\rho(c)] u_y g}}_{\epsilon_{inj}}, 
\end{equation}
where $\overline{\bm{F^\sigma} \cdot \bm{u}} = \partial_t E_\sigma = \partial_t \int (\sigma ds)$ is the surface energy and $ds$ is the surface element\cite{joseph_1976}.
It is easy to verify from Table~\ref{tab:runs} that  in the statistically steady state 
$\epsilon_{inj} \approx \epsilon_{\mu}$. It is important to note that the presence of spurious currents at bubble interface leads to additional numerical dissipation in VOF as well as front-tracking method \cite{popinet1999,popinet2018}. These effects are typically severe at large $\Ga$ and $\At$ \cite{inno20}. Therefore, even for moderate $\Ga$ and $\At$, we use high grid resolution and observe reasonable agreement between steady state values of $\epsilon_{\mu}$ and $\epsilon_{inj}$ (see Table~\ref{tab:runs}).

\subsubsection{Energy dissipated by the wakes}

\begin{figure}[!h]
    \centering
\includegraphics[width=0.9\linewidth]{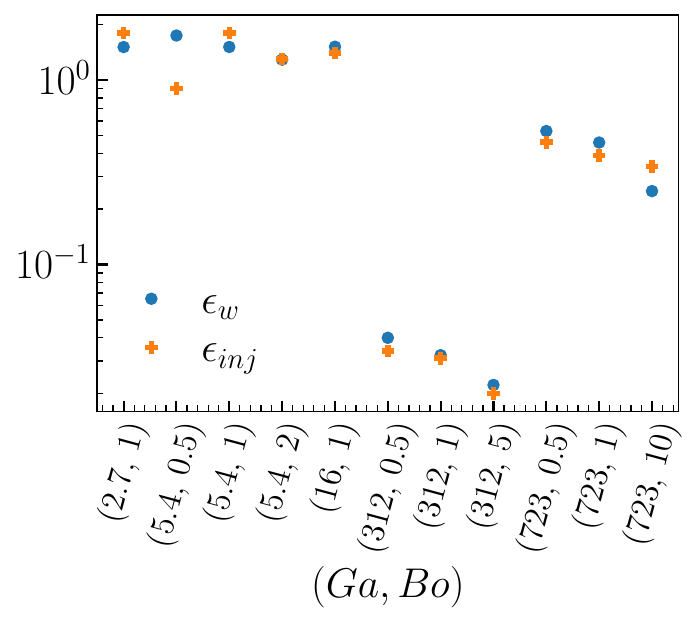}
\caption{\label{fig:disswakes} Comparison of the energy injection rate $\epsilon_{inj}$ and the estimation of the dissipation rate due to the bubble wakes \eqref{eq:wake} \cite{lance_1991}  for different $(\Ga,\Bo)$ [runs {\tt NB1-NB12}].}
\end{figure}

As the bubbles rise within a swarm, the interaction of wakes leads to psuedo-turbulence. The energy dissipated by the wakes can be estimated  as \cite{lance_1991}
\begin{equation}
\epsilon_w \sim C_D \phi \rho_f (\delta \rho g \mathcal{D}/\rho_f)^{3/2}/\mathcal{D}, 
\label{eq:wake}
\end{equation}
where $C_D$ is the drag coefficient. Assuming $\epsilon_w$ to be the dominant dissipation
mechanism, we expect it to be comparable to both the viscous dissipation $\epsilon_\nu$ and the
energy injected by buoyancy $\epsilon_{inj}$.  Although it is difficult to estimate the bubble
suspension's drag coefficient, we find $\epsilon_{inj} \approx \epsilon_w$ with $C_D=1/2$ (see \Fig{fig:disswakes} and Table~\ref{tab:runs}).

\subsection{Energy spectrum and scale-by-scale energy budget \label{esp}}
The energy spectrum $E^{uu}_k$ and the co-spectrum $E^{\rho uu}_k$ are defined as follows:
\begin{eqnarray}
\displaystyle E^{uu}_k &\equiv & \sum_{k-1/2<m<k+1/2} |\hat{\bm{u}}_m|^2, \nonumber \\
\displaystyle E^{\rho uu}_k &\equiv & \sum_{k-1/2<m<k+1/2} \Re[\hat{(\rho {\bm u})}_{-m} \hat{\bm{u}}_m]\equiv d \mathscr{E}_k/dk. \nonumber
\end{eqnarray}

We follow the procedure described in  \cite{pan20,frisch} and apply a low-pass filter to Eq.~\eqref{eqn:mom} to obtain the following energy budget equation   
\begin{equation}
   \partial_t\mathscr{E}_k = -\Pi_k -  \mathscr{P}_k + \mathscr{D}_k  + \mathscr{F}^\sigma_k + \mathscr{F}^g_k,    
   \label{eq:cumue}
\end{equation}   
with
\begin{eqnarray}
\nonumber
\mathscr{E}_k &=& \frac{1}{2}\overline{ \bm{u}^<_k\cdot(\rho\bm{u})^<_k}, \\
\nonumber
\Pi_k &=& \frac{1}{2}\overline{(\rho\bm{u})^<_k \bm{\cdot} (\bm{u}\bm{\cdot}\nabla\bm{u})^<_k} +
\overline{\bm{u}^<_k\bm{\cdot} (\bm{u}\bm{\cdot}\nabla\rho\bm{u})^<_k},\\
\nonumber
\mathscr{D}_k&=& \frac{1}{2}\overline{(\rho\bm{u})^<_k\bm{\cdot}\left(\nabla\bm{\cdot} [ 2\mu {\cal S}]/\rho\right)^<_k} + \overline{\bm{u}^<_k\bm{\cdot}(\nabla \bm{\cdot} [2 \mu {\cal S}])^<_k},\\
\nonumber
\mathscr{F}^\gamma_k &=& \frac{1}{2}\overline{(\rho\bm{u})^<_k\bm{\cdot}\left(\bm{F}^\gamma / \rho \right)^<_k} + \overline{\bm{u}^<_k\bm{\cdot}(\bm{F}^\gamma)^<_k},~\rm{and}\\
\nonumber
\mathscr{P}_k &=& \frac{1}{2}\overline{(\rho\bm{u})^<_k \bm{\cdot} \left({\nabla p}/{\rho}\right)^<_k}.
\end{eqnarray}
Here, $\mathscr{E}_k$ is
the cumulative energy  up to wave-number $k$,  $\Pi_k$ is the energy flux through wave-number $k$,  $\mathscr{D}_k$ is the cumulative energy dissipated, the contribution due to surface tension and buoyancy forces  is $2\mathscr{F}^\gamma_k$ ($\gamma$ represents either $\sigma$ or $g$). In crucial departure from the uniform density flows, we find a non-zero  cumulative pressure contribution $2\mathscr{P}_k$. The $<$ superscript above indicates low-pass filter upto wave-number $k$.  Note that in the Boussinesq regime, the density field is uniform, i.e., $\rho=\rho_a$ and ${\mathscr P}_k=0$ \cite{pan20}.

\subsubsection{Low $\Ga$, High $\At$ \rm{(}$\tt{NB1} - \tt{NB5}$\rm{)}} 
\begin{figure}[!ht] 
\begin{center}
  \includegraphics[width=0.9\linewidth]{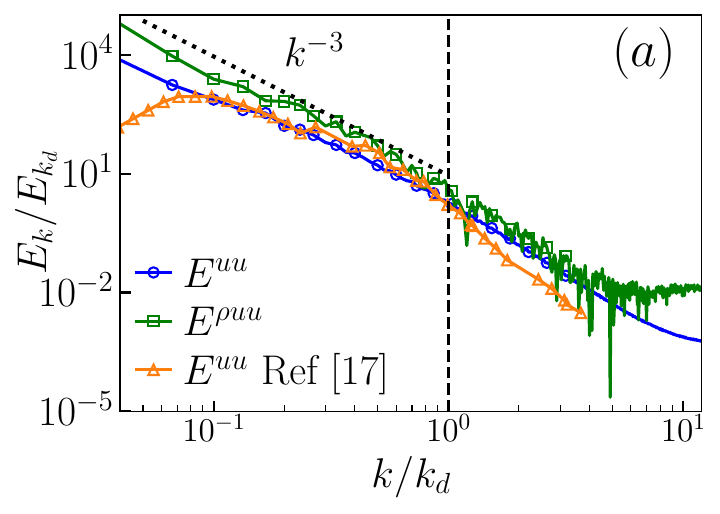}\\
  \includegraphics[width=0.9\linewidth]{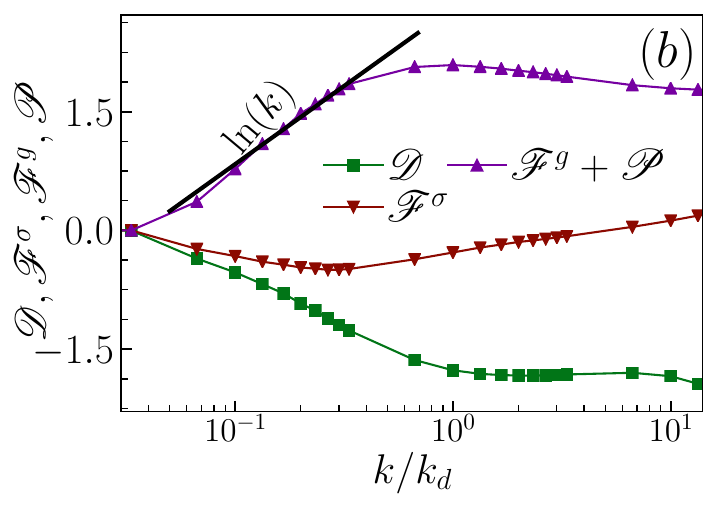}\\
  {\includegraphics[width=0.9\linewidth]{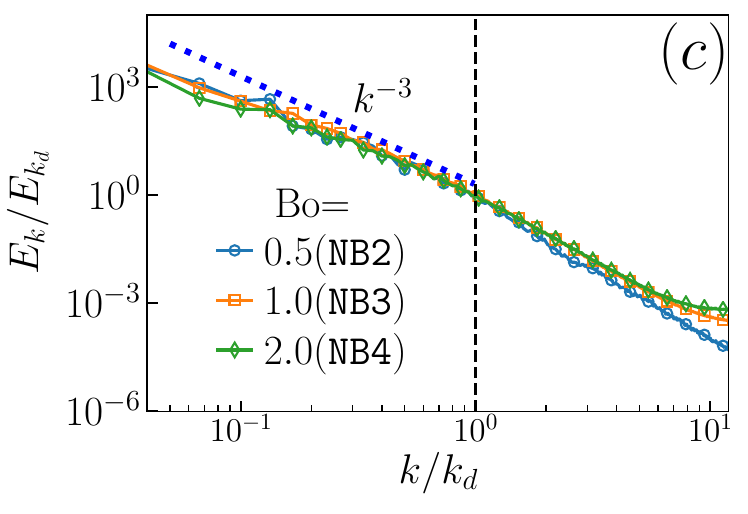}}
        \end{center}
  \caption{\label{plot:tryg}  (a) Log-log plot of spectrum and co-spectrum for $\tt{NB3}$.
  Vertical dotted line represents $k=k_d$. (b) Semi-log plot of cumulative contribution due to
  viscosity $\mathscr{D}_k$, net energy injection $\mathscr{F}_k^g + \mathscr{P}_k$, and surface
  tension $\mathscr{F}_k^\sigma$. Least squares fit  $(\mathscr{F}^g + \mathscr{P})_k=0.85\ln(27.3k/k_d)$ [black line,  $0.05\leq k/k_d \leq 0.5$].  (c) Log-log plot of spectrum $E^{uu}_k$  versus $k/k_d$ for high At runs $\tt {NB2}$-$\tt {NB4}$. }
  \end{figure}

Initial studies in two-dimensions used the front-tracking algorithm and investigated
buoyancy-driven bubbly flows at low $\Ga$~\cite{Tryg1996,esmaeeli_1998,esmaeeli_1999}. Below we
compare results of our volume of fluid (VOF) simulations with identical parameters (see Table~\ref{tab:runs}) as the $144$-bubble front-tracking simulation in  Ref.~\cite{Tryg1996}. Note that in our simulations, in contrast to  \cite{Tryg1996}, the coalescence and breakup of the bubbles is allowed.  

The plot of the kinetic energy spectrum  $E^{uu}_k$ and the co-spectrum  $E^{\rho uu}_k$ (see
\subfig{plot:tryg}{a})  show the presence of a $k^{-3}$ scaling for $k<k_d$. The energy spectrum obtained from our DNS and
\cite{Tryg1996} are in good agreement. The scale-by-scale energy budget analysis reveals that
for $k<k_d$, dominant balance is between the net energy production and viscous dissipation.
Assuming energy production to only depend on $\epsilon_{inj}$ and $k$~\cite{lance_1991}, we
expect the cumulative energy production $(\mathscr{F}^g + \mathscr{P})_k \sim \epsilon_{inj}
\ln(k)$. Consistent with the predicted scaling,  a least square fit in the range $0.05  \leq
k/k_d \leq 0.5$ gives $(\mathscr{F}^g + \mathscr{P})_k = 0.85 \ln(27.3 k/k_d)$ (see
\subfig{plot:tryg}{b}). The balance of net production $d(\mathscr{F}^g + \mathscr{P})_k/dk$   
with viscous dissipation $\nu k^2 E^{uu}_k$  [\subfig{plot:tryg}{b}] explains the observed
scaling $E^{uu}_k \sim k^{-3}$ for $k\leq k_d$. The contribution due to surface tension
${\mathscr F}^\sigma$ is negligible in this regime. In \subfig{plot:tryg}{c} we show that the
scaling of the energy spectrum is insensitive to variations in $\Bo=0.5-2$.

\subsubsection{High $\Ga$, High $\At$ \rm{(}$\tt{NB10-NB12}$\rm{)}}
\begin{figure}[!h] 
    \begin{center}
        {\includegraphics[width=0.9\linewidth]{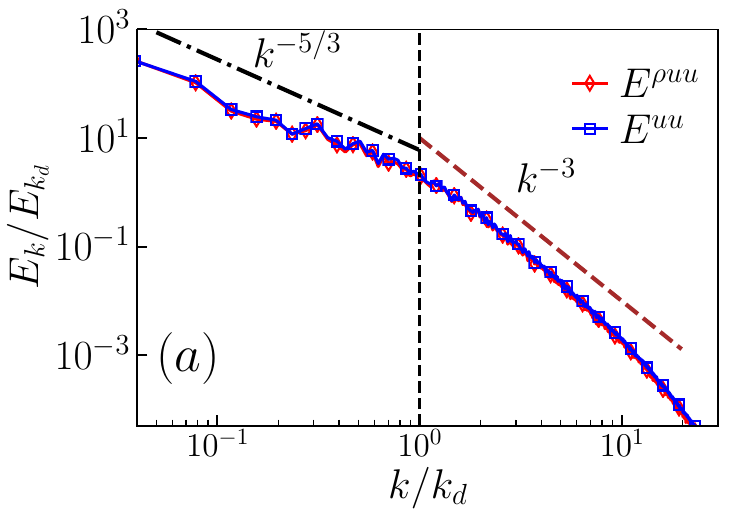}}\\
        {\includegraphics[width=0.9\linewidth]{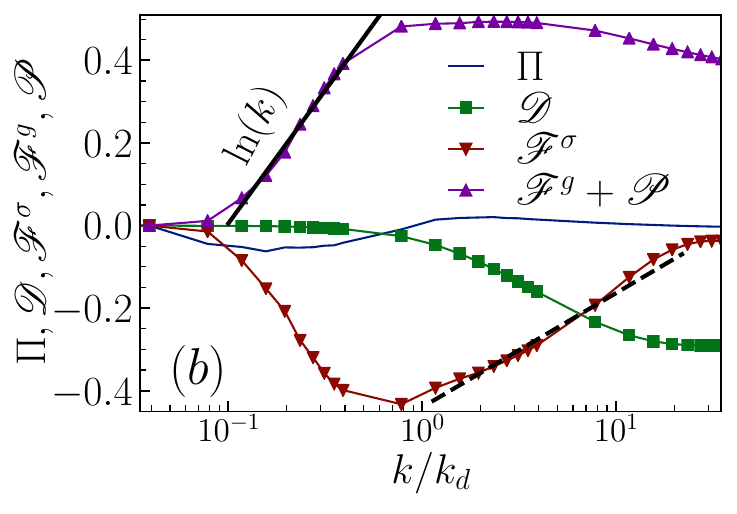}}\\
        {\includegraphics[width=0.9\linewidth]{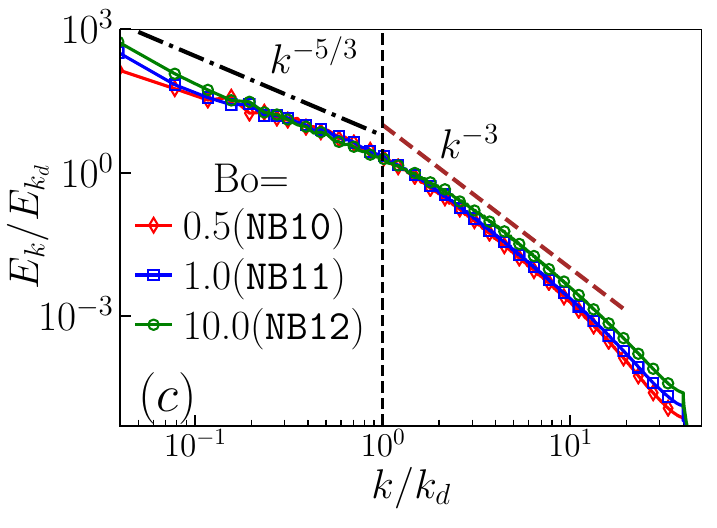}}
    \end{center}
    \caption{\label{plot:ca_hat} (a) Log-log plot of the spectrum $E^{uu}_k$ and the co-spectrum
    $E^{\rho u u}_k$ versus $k/k_d$ for high $\Ga$, high $\At$ run $\tt {NB11}$. The dash-dot
line indicates the $k^{-5/3}$ scaling  whereas, the dashed line indicates $k^{-3}$ scaling. The
wave-number $k=k_d$ is shown by a vertical dotted line. (b) Cumulative contribution of viscous
dissipation $\mathscr{D}_k$, energy flux $\Pi_k$, net cumulative energy injected $(\mathscr{F}^g + \mathscr{P})_k$, the surface tension contribution $\mathscr{F}_k^\sigma$. Least squares fits  $(\mathscr{F}^g + \mathscr{P})_k=0.28\ln(10.2k/k_d)$ (black dashed line,  $0.12\leq k/k_d \leq 0.6$), and $\mathscr{F}_k^{\sigma}=0.12\ln(0.025k/k_d)$ (black dashed line,  $1.1\leq k/k_d \leq 15.0$). (c) Log-log plot of spectrum $E^{uu}_k$  versus $k/k_d$ for high At runs $\tt {NB10}$ - $\tt {NB12}$.}
\end{figure}

On increasing the $\Ga$ number we find that the energy budget is dramatically altered.  The
spectrum $E^{uu}_k$ and the co-spectrum $E^{\rho u u}_k$ (\subfig{plot:ca_hat}{a}) show 
$k^{-5/3}$ scaling for $k<k_d$ and $k^{-3}$ scaling for $k>k_d$. In \subfig{plot:ca_hat}{b} we
plot different contributions from the scale-by-scale energy budget equation. For $k<k_d$, the
net energy injected ${{\mathscr F}^g+P}\sim \epsilon_{inj} \ln(k)$ is partly absorbed by surface
tension and, similar to inverse energy cascade in fluid turbulence,  we also  find a negative
energy flux  $\Pi_k$ for $k<k_d$. Not surprisingly, therefore, the energy spectrum 
$E^{uu}_k\sim k^{-5/3}$~\cite{frisch} for $k<k_d$.   For $k>k_d$, the energy absorbed by surface
tension is redistributed to small-scales. Assuming the cumulative surface tension contribution
to only depend on $\epsilon_w$ and $k$, we expect ${\mathscr F}^{\sigma}_k \sim \epsilon_{w}
\ln(k)$ [\subfig{plot:ca_hat}{b}]. A least square fit to  $\mathscr{F}^\sigma_k$  confirms the
logarithmic scaling. Finally, the balance  viscous dissipation $\nu k^2 E_k^{uu}$  with  the
energy transfer because of the surface tension $d\mathscr{F}^\sigma/dk \sim k^{-1}$ explains the
observed $E_k^{uu}\sim k^{-3}$ scaling. This balance for $k>k_d$ is similar to what has been
observed in 3D pseudo-turbulence \cite{pan20}. Finally, in \subfig{plot:ca_hat}{c} we show that
similar to the low $\Ga$, the scaling of the energy spectrum is insensitive to the changes in
the $\Bo$.

\subsubsection{ High $\Ga$, Low $\At$ \rm{(}{$\tt {NB6}-\tt{NB8}$}\rm{)}} 
\begin{figure}
\begin{center}
  {\includegraphics[width=\linewidth]{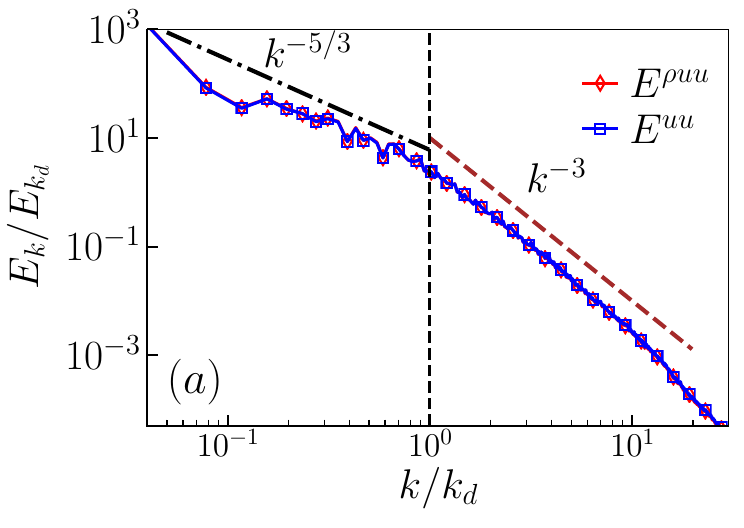}}\\
  {\includegraphics[width=\linewidth]{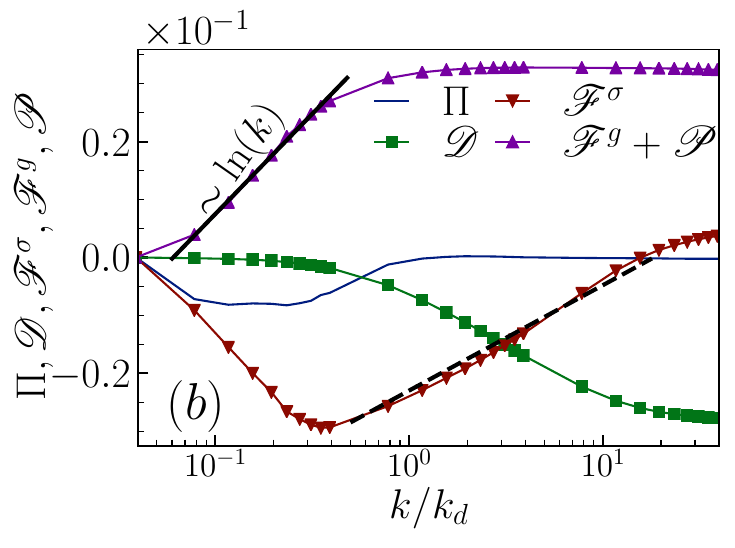}}
   \end{center}
  \caption{\label{plot:ca_hgalat} (a) Log-log plot of the spectrum $E^{uu}_k$ and the
      co-spectrum $E^{\rho u u}_k$ versus $k/k_d$ for high $\Ga$, low $\At$ run $\tt {NB7}$. The
      dash-dot line indicates the $k^{-5/3}$ scaling  whereas, the dashed line indicates
      $k^{-3}$ scaling. The wave-number $k=k_d$ is shown by a vertical dotted line.  (b) The energy flux $\Pi_k$, Cumulative contribution of viscous dissipation $\mathscr{D}_k$, net cumulative energy injected
      $\mathscr{F}_k^g + \mathscr{P}_k$, the surface tension contribution
      $\mathscr{F}_k^\sigma$. Least squares fits $(\mathscr{F}^g + \mathscr{P})_k=0.015\ln(16.4k/k_d)$ (black line,$0.07\leq k/k_d \leq 0.6$), and $\mathscr{F}_k^{\sigma}=0.008\ln(0.06k/k_d)$ (black dashed line, $0.6\leq k/k_d \leq 15.0$).} 
\end{figure}

In the earlier section, we presented the results for pseudo-turbulence in the high $\Ga$, high $\At$ number regime and showed that the statistical properties of PT are robust to changes in the Bond number. In this section, we show that the phenomenology of PT remains the same even for low $\At=0.08$ number.  In \subfig{plot:ca_hgalat}{a} we plot the spectrum for low $\At=0.08$ and show, similar to high $\At$ runs, $E_k^{uu}\sim k^{-5/3}$ for $k<k_d$ and   $E_k^{uu}\sim k^{-3}$ for $k>k_d$  \footnote{We have also verified (not shown) that the scaling of the energy spectrum for $\At=0.08$ does not depend on the $\Bo$.}. From our energy budget analysis we make the following observations (see \subfig{plot:ca_hgalat}{b}):  (i) For $k<k_d$  a non-zero energy flux $\Pi_k$;  (ii) For $k>k_d$, $\Pi_k \approx 0$ and the energy injected by surface tension 
($|\mathscr{F}^\sigma_k|$ increases)-because of bubble shape undulations-  is balanced  by viscous dissipation ($\mathscr{D}_k$  decreases).  A non-zero $\Pi_k$ indicates  presence of an inverse energy cascade and hence, using Kolmogorov's phenomenology \cite{frisch}, $E^{uu}_k\sim k^{-5/3}$. On the other hand for $k>k_d$, the balance of energy redistributed by surface tension $\sim \epsilon_w/k$  with viscous dissipation gives $E^{uu}_k\sim k^{-3}$ . 

\begin{figure}
\begin{center}
  {\includegraphics[width=0.49\linewidth]{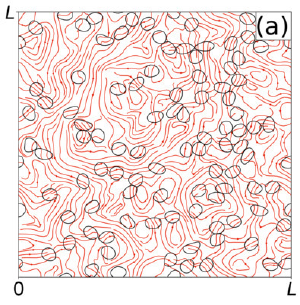}}
  {\includegraphics[width=0.49\linewidth]{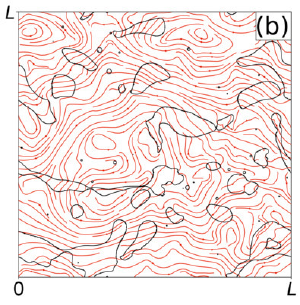}}\\
  {\includegraphics[width=\linewidth]{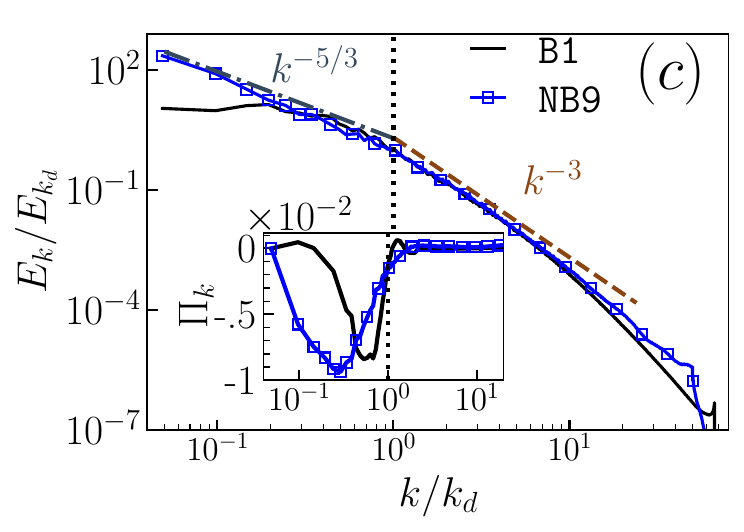}} \\
  {\includegraphics[width=\linewidth]{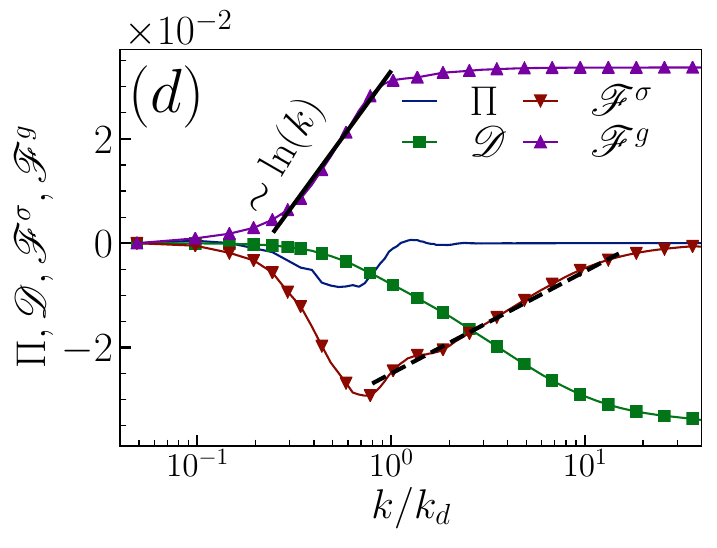}}\\
  \end{center}
  \caption{\label{plot:bou_nb}(a,b) Representative steady-state snapshot of the  velocity streamline with overlaid bubble positions for $\tt{B1}$ (front-tracking) and $\tt{NB9}$ (VOF) runs. (c) Log-log plot of
     energy spectra $E^{uu}_k$ versus $k/k_d$ for run $\tt{B1}$ and $\tt{NB9}$. (Inset) Semilog
     plot showing negative energy flux $\Pi_k$ versus $k/k_d$. Vertical dashed line corresponds
     to $k = k_d$. (d) Cumulative contribution of viscous dissipation $\mathscr{D}_k$, energy flux $\Pi_k$, 
     energy injected because of buoyancy $\mathscr{F}^g_k$, and the surface tension contribution 
     $\mathscr{F}^\sigma_k$ versus $k/k_d$ for $\tt{B1}$. Least squares fits  $\mathscr{F}^g_k=0.022\ln(4.3k/k_d)$(black line, $0.2\leq k/k_d \leq 1.5$), and $\mathscr{F}_k^{\sigma}=0.0085\ln(0.05k/k_d)$ (black dashed line, $1 \leq k/k_d \leq 15.0$. For $k<k_d$, $d\mathscr{F}_k^\sigma/dk$ balances $d\mathscr{F}^g_k/dk$ whereas, for $k>k_d$, $d\mathscr{F}_k^\sigma/dk$ is balanced by viscous dissipation $d\mathscr{D}_k/dk$. } 
\end{figure}

\subsubsection{Effect of merger and breakup \rm{(}$\tt{B1}$, {$\tt {NB9}$}\rm{)}}
To further highlight the robustness of the energy spectrum, we conduct  DNS using a front-tracking method~\cite{Tryg2001}  where breakup and coalescence of bubbles is not allowed.  We refer the reader to \cite{pan20} for details of the front-tracking scheme. We plot the snapshot of the bubble positions overlaid on the corresponding velocity streamlines for the front-tracking run $\tt{B1}$ and the bubble suspension from VOF run $\tt{NB9}$ in \subfig{plot:bou_nb}{a,b}. We find the bubbles are significantly deformed and their shape is nearly ellipsoidal. During the evolution, the average bubble diameter in $\tt{NB9}$ run remains close to the initial  diameter. The energy spectrum obtained from our 2D  runs ($\tt {B1}$ and $\tt{NB9}$) are in excellent agreement, $E^{uu}_k\sim k^{-5/3}$ for $k<k_d$ and $E^{uu}_k\sim k^{-3}$ for $k>k_d$  [\subfig{plot:bou_nb}{c}]. Note that the region of negative energy flux (and $k^{-5/3}$ scaling) is broader for $\tt{NB9}$ because coalescence and breakup leads to a bubble size distribution and an enhanced injection because of larger bubbles. Thus, coalescence of bubbles does not alter the scaling behaviour.

\subsubsection{Pseudo-turbulence in 2D versus 3D}
We now contrast the pseudo-turbulence spectral balances in 2D with our recent study in 3D
\cite{pan20}.  In both cases, buoyancy injects energy at scales comparable to the bubble
diameter. In 3D, the energy transfer due to the surface tension and forward kinetic energy flux
balances viscous dissipation leading to the $k^{-3}$ scaling for scales smaller than the bubble
diameter. In contrast, in 2D, we show an inverse energy cascade from the bubble diameter
scale to larger scales. Only the surface tension contribution transfers the energy to scales smaller than the bubble diameter. The viscous dissipation balances energy transfer by surface tension leading to the $k^{-3}$ scaling.

\section{Conclusion\label{concl}}
To conclude, we have investigated the spectral properties of buoyancy driven 
 bubbly flows. Using scale-by-scale energy budget we show that  a non-zero 
negative energy flux in two-dimension that is indicative of an inverse cascade and 
leads to a $k^{-5/3}$ spectrum for scales larger than the bubble diameter. Although flow around an individual bubble 
strongly depends on the $\At$  \cite{tri15,bha81,wang2014experimental,piedra_2015}, intriguingly, the scaling
that we observe is not sensitive to the density contrast ($\At$). Our scale-by-scale budget analysis reveals that in two-dimensional bubbly flows the $k^{-3}$ scaling observed at large $\Ga$ is because of a balance between energy production due to surface tension and viscous dissipation.
\\ \\
R.R. conducted VOF simulations and V.P. conducted FT simulations. All authors analysed the results and reviewed the manuscript. 

 \begin{acknowledgement}
We thank D. Mitra and S. Banerjee for discussions, support from intramural funds at TIFR Hyderabad from the Department of Atomic Energy (DAE), India and DST (India) Project No. ECR/2018/001135.
\end{acknowledgement}

\bibliographystyle{epj}
\bibliography{2Dbubblyflowspaper}

\end{document}